\documentclass[conference,a4paper]{IEEEtran}
\usepackage[utf8]{inputenc}
\usepackage[T1]{fontenc}
\usepackage{url}
\usepackage{ifthen}
\usepackage{cite}
\usepackage[cmex10]{amsmath} % Use the [cmex10] option to ensure complicance
\usepackage{graphicx}
\usepackage{subfigure}
\usepackage{algorithm}
\usepackage{algorithmic}
\usepackage{color,xcolor}

\definecolor{red}{rgb}{1,0,0}
\usepackage{color,soul}
\definecolor{lightblue}{rgb}{.90,.95,1}
\sethlcolor{yellow}

\begin{document}
\title{Combating Error Propagation in Window Decoding of Braided Convolutional Codes}

% %%% Single author, or several authors with same affiliation:
% \author{%
%   \IEEEauthorblockN{Stefan M.~Moser}
%   \IEEEauthorblockA{ETH Zürich\\
%                     ISI (D-ITET)\\
%                     CH-8092 Zürich, Switzerland\\
%                     Email: moser@isi.ee.ethz.ch}
% }

%%% Several authors with up to three affiliations:
%\author{%
%  \IEEEauthorblockN{Stefan M.~Moser}
%  \IEEEauthorblockA{ETH Zürich\\
%                    ISI (D-ITET), ETH Zentrum\\
%                    CH-8092 Zürich, Switzerland\\
%                    Email: moser@isi.ee.ethz.ch}
%  \and
%  \IEEEauthorblockN{Albus Dumbledore and Harry Potter}
%  \IEEEauthorblockA{Hogwarts School of Witchcraft and Wizardry\\
%                    Hogwarts Castle\\
%                    1714 Hogsmeade, Scotland\\
%                    Email: \{dumbledore, potter\}@hogwarts.edu}
%}

%%% Many authors with many affiliations:
 \author{%
   \IEEEauthorblockN{Min Zhu\IEEEauthorrefmark{1},
                     David G.~M. Mitchell\IEEEauthorrefmark{2},
                     Michael Lentmaier\IEEEauthorrefmark{3},
                     Daniel J. Costello, Jr.\IEEEauthorrefmark{4},
                     and Baoming Bai\IEEEauthorrefmark{1}}
   \IEEEauthorblockA{\IEEEauthorrefmark{1}%
                     State Key Laboratory of ISN, Xidian University, Xi'an, P. R. China,
                     mzhu@xidian.edu.cn, bmbai@mail.xidian.edu.cn}
   \IEEEauthorblockA{\IEEEauthorrefmark{2}%
                     Klipsch School of Electrical and Computer Engineering, New Mexico State University, Las Cruces, NM , USA,
                     dgmm@nmsu.edu}
   \IEEEauthorblockA{\IEEEauthorrefmark{3}%
                     Department of Electrical Engineering, University of Notre Dame, Notre Dame, IN, USA,
                     dcostel1@nd.edu}
   \IEEEauthorblockA{\IEEEauthorrefmark{4}%
                     Department of Electrical and Information Technology, Lund University, Lund, Sweden,
                     michael.lentmaier@eit.lth.se}
 }

\maketitle

%%%%%%
%% Abstract:
%% If your paper is eligible for the student paper award, please add
%% the comment "THIS PAPER IS ELIGIBLE FOR THE STUDENT PAPER
%% AWARD." as a first line in the abstract.
%% For the final version of the accepted paper, please do not forget
%% to remove this comment!
%%
\begin{abstract}
  In this paper, we study sliding window decoding of braided convolutional codes (BCCs) in the context of a streaming application, where decoder error propagation can be a serious problem. A window extension algorithm and a resynchronization mechanism are introduced to mitigate the effect of error propagation. In addition, we introduce a soft bit-error-rate stopping rule to reduce computational complexity, and the tradeoff between performance and complexity is examined. Simulation results show that, using the proposed window extension algorithm and resynchronization mechanism, the error performance of BCCs can be improved by up to three orders of magnitude with reduced computational complexity.
\end{abstract}

\section{Introduction}
Braided convolutional codes, first introduced in \cite{WeizhangIT2010}, are a counterpart to braided block codes (BBCs)~\cite{FeltstromIT2009} which can be regarded as a diagonalized version of product codes \cite{Elias1954} or expander codes \cite{Sipser1996}. In contrast to BBCs, BCCs use short constraint length convolutional codes as component codes. The encoding of BCCs can be described by a two-dimensional sliding array, where each symbol is protected by two component convolutional codes. BCCs are a type of parallel-concatenated convolutional code in which the parity outputs of one component encoder are fed back and used as inputs to the other component encoder at the succeeding time unit. Two variants of BCCs were considered in \cite{WeizhangIT2010}. Tightly braided convolutional codes (TBCCs) are obtained if a dense array is used to store the information and parity symbols. This construction is deterministic and simple but performs relatively poorly due to the absence of randomness. Alternatively, sparsely braided convolutional codes (SBCCs) that employ random permutors have low density, resulting in improved iterative decoding performance \cite{WeizhangIT2010}. Moloudi \textit{et al}. considered SBCCs as spatially coupled turbo-like codes and showed that threshold saturation occurs for SBCCs over the binary erasure channel \cite{Saeedeh2017IT,Saeedeh2016ISITA}. SBCCs can operate in bitwise or blockwise modes, according to whether convolutional or block permutors are employed. It was also shown numerically that the free (minimum) distance of bitwise (blockwise) SBCCs grows linearly with the overall constraint length, leading to the conjecture that SBCCs are asymptotically good \cite{WeizhangIT2010,Saeedeh2016ISITA}.

Due to their turbo-like structure, BCCs can be decoded with iterative decoding. Analogous to LDPC convolutional codes \cite{LentmaierTIT2010, IyengarTIT2012}, SBCCs can employ sliding window decoding for low latency operation. Unlike window decoding of LDPC convolutional codes, which typically uses an iterative belief-propagation (BP) message passing algorithm, window decoding of SBCCs is based on the Bahl-Cocke-Jelinek-Raviv (BCJR) algorithm. It has been shown that blockwise SBCCs with sliding window decoding have capacity-approaching performance \cite{Zhu2017TCOM}, but for large frame lengths or streaming applications, SBCCs can suffer from decoder error propagation. That is, once a block decoding error occurs, decoding of the following blocks is affected, which can cause a continuous string of block errors, resulting in unacceptable performance loss.

In this paper, we study several error propagation mitigation techniques for SBCCs. Specifically, a window extension algorithm and a resynchronization mechanism are introduced to combat error propagation. In addition, a soft bit-error-rate (BER) stopping rule is proposed to reduce decoding complexity and, the resulting tradeoff between  decoding performance and decoding complexity is explored.
%Following a review of SBC codes in Section II and the introduction of error propagation mitigation in Section III, we examine the error performance of SBC codes with proposed algorithms during window decoding in Section IV. Concluding remarks are given in Section V.

\section{Continuous Transmission of Braided Convolutional Codes}
In this section, we briefly review continuous encoding and sliding window decoding of blockwise SBCCs. For details, please refer to \cite{WeizhangIT2010} and \cite{Zhu2017TCOM}.

\subsection{Continuous Encoding}
\begin{figure*}[t]
        \center
        \includegraphics[width= 0.8 \textwidth]{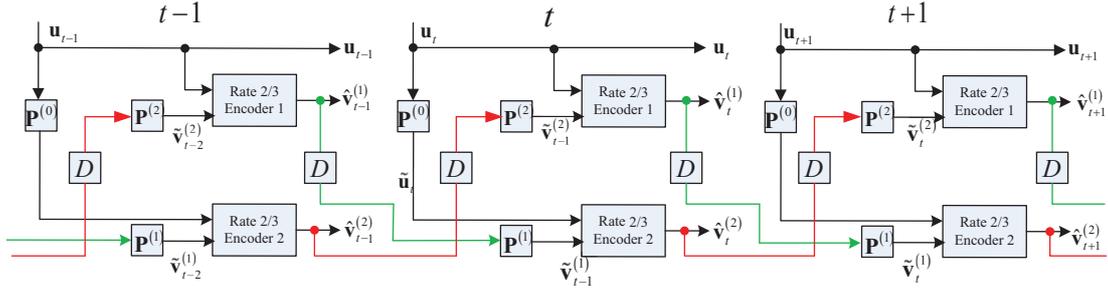}\\
        \caption{Continuous encoder chain for a rate $R = 1/3$ blockwise SBCC.}
        \label{fig:Chain_Encoder}
    \end{figure*}

Sparsely braided convolutional codes are constructed using an infinite two-dimensional array consisting of one horizontal and one vertical encoder. These two encoders are linked through parity feedback. In this manner, the systematic and parity symbols are ``braided'' together. In this paper, we limit ourselves to rate $R = 1/3$ blockwise SBCCs as an example. In this case, the information sequence enters the encoder in a block-by-block manner with a relatively large block size. Fig. \ref{fig:Chain_Encoder} is a conceptual illustration of the continuous encoding process for a rate $R = 1/3$ blockwise SBCC, which utilizes two recursive systematic convolutional (RSC) component encoders each of rate ${R_{cc}} = 2/3$, where ${{\bf{P}}^{\left( 0 \right)}}$, ${{\bf{P}}^{\left( 1 \right)}}$, and ${{\bf{P}}^{\left( 2 \right)}}$ are each block permutors of length $T$. The information sequence is divided into blocks of length $T$ symbols, i.e., ${\bf{u}} = \left( {{{\bf{u}}_0},{{\bf{u}}_1}, \ldots ,{{\bf{u}}_t}, \ldots } \right)$, where ${{\bf{u}}_t} = \left( {{u_{t,1}},{u_{t,2}}, \ldots ,{u_{t,T}}} \right)$, at time $t$ ${{\bf{u}}_t}$ is interleaved using ${{\bf{P}}^{\left( 0 \right)}}$ to form ${{\bf{\tilde u}}_t}$, and ${{\bf{u}}_t}$ and ${{\bf{\tilde u}}_t}$ enter the component encoders. The parity outputs $\hat {\bf{v}}_{t}^{\left( i \right)}$, $i \in \{ 1,2\} $, at time $t$ are delayed by one time unit, interleaved using ${{\bf{P}}^{\left( 1 \right)}}$, and ${{\bf{P}}^{\left( 2 \right)}}$, and then enter the component encoders as the input sequences $\tilde {\bf{v}}_{ t}^{\left( i \right)}$, $i \in \{ 1,2\} $, at time $t+1$. The information sequence ${\bf{u}}$, the parity output sequence ${\bf{\hat v}}_t^{\left( 1 \right)}$ of encoder 1, and the parity output sequence ${\bf{\hat v}}_t^{\left( 2 \right)}$ of encoder 2 are sent over the channel.
%    \begin{figure}\center
%        % Requires \usepackage{graphicx}
%        \includegraphics[width= 0.45 \textwidth]{BCC_encoder.eps}\\
%        \caption{Encoder for a rate $R = 1/3$ blockwise SBC code.}
%        \label{fig:Encoder}
%    \end{figure}
%In order to depict the continuous encoding process \hl{conceptually}, a chain of encoders that operates at different time instants is illustrated in Fig. \ref{fig:Chain_Encoder}. At each time instant, there is a turbo-like encoder which consists of \hl{two} parallel concatenated RSC component encoders. These turbo-like encoders are coupled by feeding the parity sequence generated at \hl{the} current time instant to the encoders at \hl{the} next time instant\hl{, so that} the coupling memory is 1 in this case.
For initialization, at time instant 0, we assume that ${\bf{\tilde v}}_{ - 1}^{\left( 1 \right)} = {\bf{0}}$ and ${\bf{\tilde v}}_{ - 1}^{\left( 2 \right)} = {\bf{0}}$.
Transmission can be terminated after a frame consisting of $L$ blocks, resulting in a slight rate loss, or proceed in an unterminated (streaming) fashion, in which case the rate is given by
\begin{equation}
 R = \frac{1}{3}.
\end{equation}

\subsection{Sliding Window Decoding}

%There are the pipline decoder \cite{WeizhangISIT2005,WeizhangIT2010} and the sliding window decoder \cite{Zhu2015ITW,Zhu2017TCOM} for the SBC code. The sliding window decoder allows low-latency operation with negligible performance loss. In this paper, we focus on the sliding window decoding for SBC codes.

\begin{figure*}[t]
        \center
        \includegraphics[width= 0.8 \textwidth]{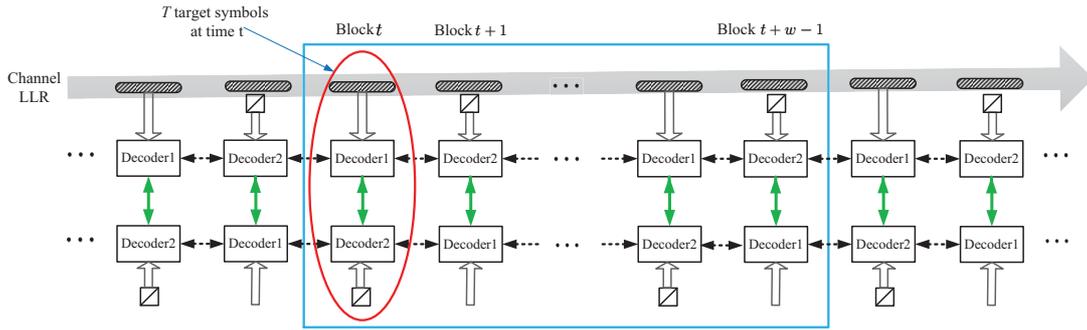}\\
        \caption{Continuous sliding window decoder for blockwise SBCCs \cite{Zhu2017TCOM}.}
        \label{fig:Decoder}
    \end{figure*}

In order to describe the proposed error propagation mitigation methods, the structure of the sliding window decoder \cite{Zhu2017TCOM} is shown in Fig. \ref{fig:Decoder}. The window size is denoted as $w $. The block at time instant $t$ is the target block for decoding in the window containing the blocks received at times $t$ to $t+w-1$. Briefly, the decoding process in a window beings with $I_1$ turbo, or \emph{vertical}, iterations on the target block at time $t$, during which the two component convolutional codes pass soft messages on the $T$ information bits in that block to each other. Then, soft messages on the parity bits are passed forward, and $I_1$ vertical iterations are performed on the block at time $t+1$. This continues until $I_1$ vertical iterations are performed on the last received block in the window. Then the process is repeated in the backward direction (from the last block to the first block in the window) with soft messages being passed back through the 2$w$ BCJR decoders. This round trip of decoding is called a \emph{horizontal} iteration. After $I_2$ horizontal iterations, the $T$ target symbols are decoded, and the window shifts forward to the next position, where the $T$ symbols at time $t+1$ become the target symbols.

\subsection{Error Propagation}
\begin{figure}[h]
        \center
        \includegraphics[width= 0.4 \textwidth]{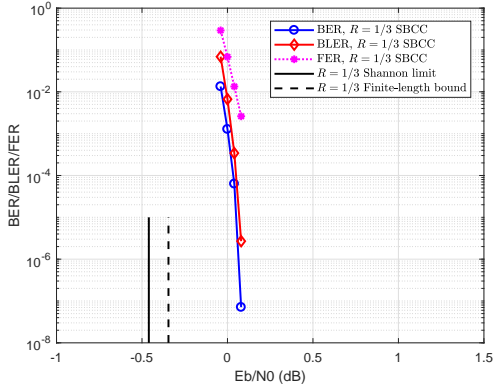}\\
        \caption{The BER, BLER, and FER performance of rate $R=1/3$ SBCCs.}
        \label{fig:Origin}

    \end{figure}

Since an encoded block in a blockwise BCC affects the encoding of the next block (see Fig. \ref{fig:Chain_Encoder}), each time a block of target symbols is decoded, the log-likelihood ratios (LLRs) associated with the decoded symbols also affect the decoding of the next block. Hence, if, after the maximum number of decoding iterations, some unreliable LLRs remain in the target block, causing a block decoding error, those unreliable LLRs can trigger a string of additional block errors, resulting in \emph{error propagation}. To illustrate this effect, we consider two identical 4-state RSC component encoders whose generator matrix is given by
\begin{equation}\label{GenerateMatrix}
{G}\left( D \right) = \left( {\begin{array}{*{20}{c}}
   1 & 0 & {\frac{1}{{1 + D + {D^2}}}}  \\
   0 & 1 & {\frac{{1 + {D^2}}}{{1 + D + {D^2}}}}  \\
\end{array}} \right).
\end{equation}
where we assume the encoders are left unterminated at the end of each block. The three block permutors ${{\bf{P}}^{\left( 0 \right)}}$, ${{\bf{P}}^{\left( 1 \right)}}$, and ${{\bf{P}}^{\left( 2 \right)}}$ were chosen randomly with the same size $T=8000$. We assume that transmission stops after a frame of $L$ blocks is decoded and a uniform decoding schedule (see \cite{Zhu2017TCOM} for details) is used. The bit error rate (BER), block error rate (BLER), and frame error rate (FER) performance for transmission over the AWGN channel with BPSK signalling is given in Fig. \ref{fig:Origin}, where the window size $w=3$, the number of vertical iteration is $I_1 = 1$, the number of horizontal iteration is $I_2 = 20$, and the frame length $L=1002$.

\begin{figure}[htbp]
   \centerline{\subfigure[One frame with different numbers of iterations, $w=3$.]{\includegraphics[width=1.8in]{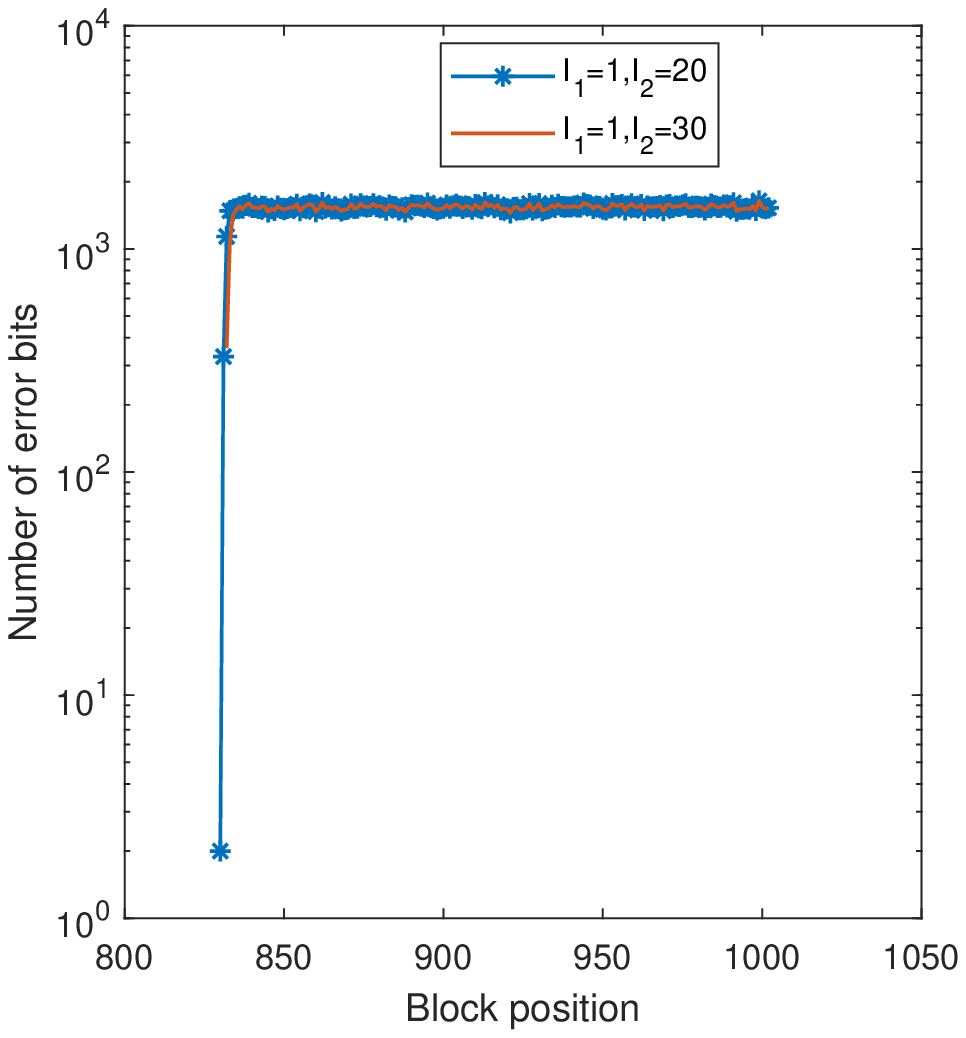}
       % where an .eps filename suffix will be assumed under latex,
       % and a .pdf suffix will be assumed for pdflatex
       \label{fig:DiffITER}}
     \hfil
     \subfigure[10000 frames with different window sizes, $I_1=1$, $I_1=20$.]{\includegraphics[width=1.8in]{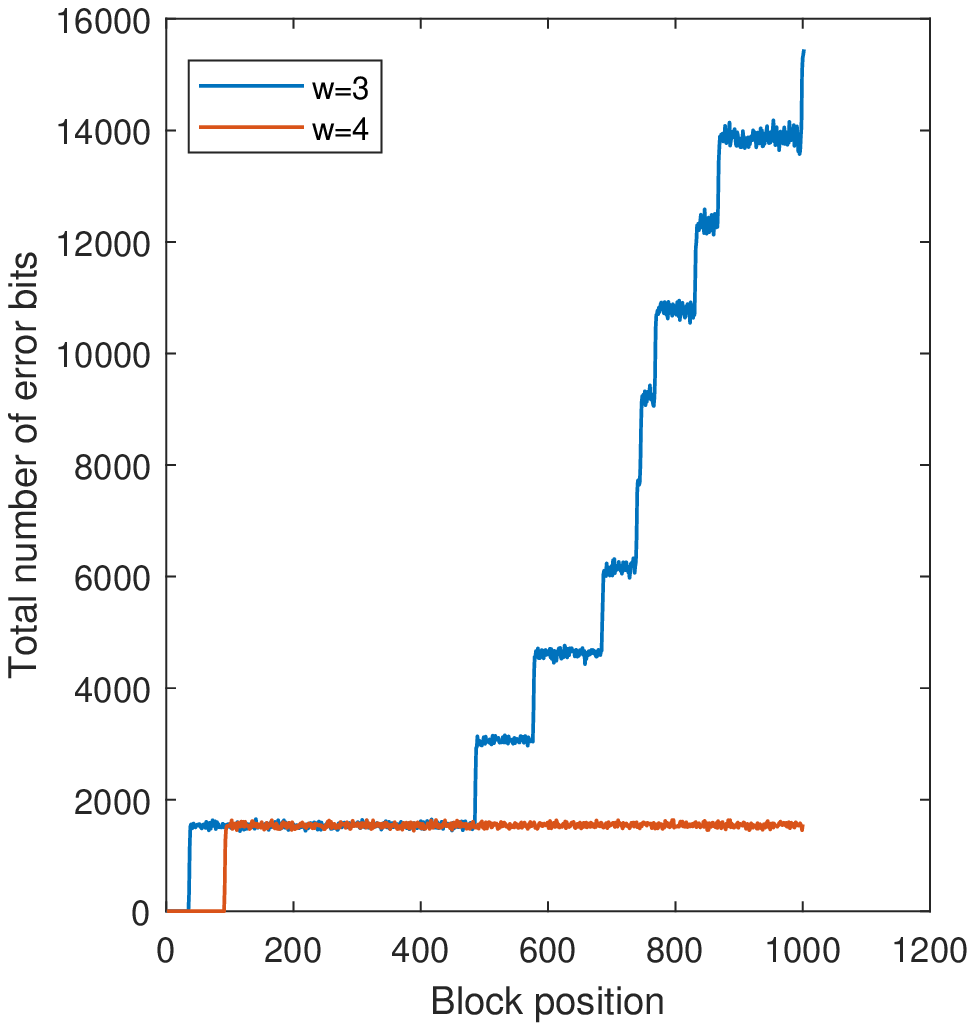}
       % where an .eps filename suffix will be assumed under latex,
       % and a .pdf suffix will be assumed for pdflatex
       \label{fig:DiffWin}}}
   \caption{The error distribution per block for rate $R=1/3$ blockwise SBCCs.}
   \label{fig:ErrDis}
 \end{figure}

From Fig. \ref{fig:Origin}, we see that the rate $R=1/3$ blockwise SBCC performs about 0.6 dB away from the Shannon limit. Even so, among the 10000 simulated frames, there were some frames that exhibited error propagation. In order to depict the error propagation phenomenon clearly, we give the bit error distribution per block of one frame with error propagation in Fig. \ref{fig:DiffITER}, for $E_b/N_0 = 0.04$ dB. We see that, for $I_2=20$, from the 830th block on, the number of error bits is large, and the errors continue to the end of the frame, a clear case of error propagation. For $I_2=30$, error propagation starts two blocks later than for $I_2=20$, but the overall effect of increasing the number of iterations is minimal.  The bit error distribution per block, based on 10000 simulated frames with two different window sizes, is shown in Fig. \ref{fig:DiffWin}, where we see that increasing the window size reduces the number of error propagation frames from 9 to 1, thus significantly improving performance.

%    \begin{figure}[h]
%        % Requires \usepackage{graphicx}
%        \center
%        \includegraphics[width= 0.45 \textwidth]{e9904_EbN004.eps}\\
%        \caption{An example of \hl{the} error distribution per block \hl{in an error propagation frame}.}
%        \label{fig:ErrDis}
%    \end{figure}

For larger frame lengths, and particularly for streaming transmission, error propagation will severely degrade the decoding performance illustrated in Fig. \ref{fig:Origin}. Hence, we now introduce two ways of mitigating the error propagation effect in sliding window decoding of SBCCs.

\section{Error Propagation Mitigation}
In this section, we propose a window extension algorithm and a resychronization mechanism to mitigate the effect of error propagation in SBCCs.

\subsection{Window Extension Algorithm}
In the paper \cite{Zhu2017TCOM} by Zhu et al., the window size is fixed during the decoding process. Based on the results presented in Fig. \ref{fig:DiffWin}, here we introduce a variable window size concept to sliding window decoding.
Before describing the window extension algorithm, we give some definitions.
Let
${\bf{L}}_d^{\left( i,j \right)} = \left\{ {l_{d,0}^{\left( i,j \right)},l_{d,1}^{\left( i,j \right)},l_{d,2}^{\left( i,j \right)}, \ldots ,l_{d,T - 1}^{\left( i,j \right)}} \right\}$ denote the decision LLRs of the $T$ information bits in the $i$th block of the current window after the $j$th horizontal iteration. Then the average absolute LLR of the $T$ information bits after the $j$th horizontal iteration is given by
$\bar L_d^{\left( {i,j} \right)} = \frac{1}{T}\sum\limits_{k = 0}^{T - 1} {\left| {l_{d,k}^{\left( i,j \right)}} \right|}$.

%
%For a binary sequence ${\bf{u}} = \left\{ {{u_0},{u_1},{u_2}, \ldots ,{u_{T - 1}}} \right\}$ of length $T$, the log-likelihood ratios (LLRs) over the binary-input additive white Gaussian noise (AWGN) channel are denoted as ${\bf{L}} = \left\{ {{l_0},{l_1},{l_2}, \ldots ,{l_{T - 1}}} \right\}$. Then the estimated BER of ${\bf{u}}$ can be given by

\begin{figure*}[ht]
        \center
    \includegraphics[width= 0.8 \textwidth]{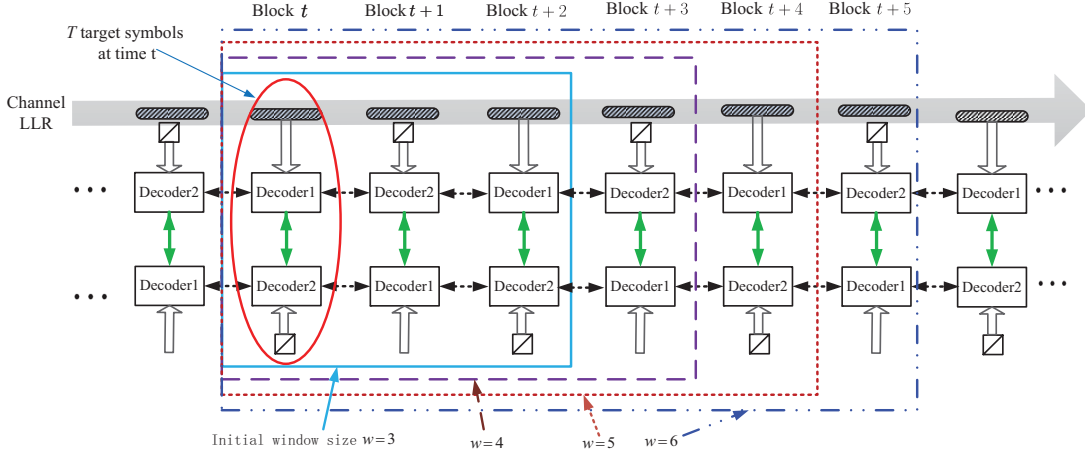}\\
    \caption{Decoder with the window extension algorithm.}
    \label{fig:WinExt}
\end{figure*}

During the decoding process, when the number of horizontal iterations reaches its maximum value $I_2$, if any of the average absolute LLRs of the first $\tau$ blocks in the current window, $1 \le \tau  \le w$, is lower than a predefined threshold $\theta$, that is, if
\begin{equation}\label{eq:2}
\bar L_d^{\left( {i,j} \right)} < {\theta}, ~~~~~i = 0,1,\ldots, \tau-1,
\end{equation}
the target block is not decoded, the window size is increased by 1, and the decoding process restarts with horizontal iteration number 0. This process continues until either the target block is decoded or the window size reaches a predefined maximum ${w_{\max }}$, in which case the target block is decoded regardless of whether \eqref{eq:2} is satisfied.
Assuming an initial window size $w=3$, Fig. \ref{fig:WinExt} illustrates that, if (\ref{eq:2}) is met, the window size increases to $w=w+1$, up to a maximum window size of ${w_{\max }}=6$.\footnote{The reason for choosing ${w_{\max }}=6$ is that, if decoding speed is not critical, we can reuse the same hardware needed to implement window size w = 3.  In other words, instead of providing additional hardware, the available w = 3 hardware can be used twice to emulate a w = 6 window in a serial, rather than parallel, implementation.} The details of the window extension process are described in Alg. 1.
%Full details of the window extension algorithm are given in an expanded version of the paper posted online [???].

%\begin{algorithm}[!t]
%\caption{Window Extension Algorithm } \label{Alg:WinExt_Alg}
%\begin{algorithmic}[1]
%\STATE Assume that the block at time $t$ is the target block in \hl{a window decoder of} size $w$ \hl{initialized with the channel LLRs of $w$ received blocks. Let $I_{count}$ denote the current number of horizontal iterations, and set $I_{count}=0$ initially. $\tau$, $\theta$, and $w_{max}$ are parameters.}
%\WHILE{$I_{count} < I_2$}
%\STATE \hl{Perform} vertical decoding and horizontal decoding;
%\STATE Every time a horizontal iteration is finished,
%\STATE $I_{count} ++$;
%\IF {$I_{count} == I_2$}
%  \STATE Calculate \hl{$\bar L_d^{\left( {i,I_2} \right)} = \frac{1}{T}\sum\limits_{k = 0}^{T - 1} {\left| {l_{d,k}^{\left( i,I_2 \right)}} \right|}$, $i=\{0,1, \ldots, \tau-1\}$.}
%  \IF{\hl{$\bar L_d^{\left( {i,I_2} \right)} < {\theta }$ for any $i \in \left\{ {0,1, \ldots ,\tau  - 1} \right\}$}}
%    \IF{$w < w_{max}$}
%     \STATE $w=w+1$.
%     \STATE $I_{count} = 0$.
%     \STATE Initialize the decoder \hl{with the channel LLRs of $w$ blocks}.
%    \ENDIF
%  \ENDIF
%\ENDIF
%\ENDWHILE
%\end{algorithmic}
%\end{algorithm}

\noindent\rule[0.0\baselineskip]{0.5\textwidth}{0.6pt}\\
$\textbf{~~~~~~~~Algorithm 1: Window Extension Algorithm}$\\
\begin{itemize}
  \item \textbf{Initialization}\\
        Assume that the block at time $t$ is the target block in a window decoder of size $w$ initialized with the channel LLRs of $w$ received blocks. Let $I_{count}$ denote the current number of horizontal iterations, and set $I_{count}=0$ initially. $\tau$, $\theta$, and $w_{max}$ are parameters.
  \item \textbf{While} $I_{count} < I_2$
        \begin{itemize}
          \item [1.] Perform vertical decoding and horizontal decoding;
          \item [2.] Every time a horizontal iteration is finished, $I_{count} ++$;
          \item [3.] \textbf{If} $I_{count} == I_2$
          \begin{itemize}
            \item [1)]Calculate~\\
            $\bar L_d^{\left( {i,I_2} \right)} = \frac{1}{T}\sum\limits_{k = 0}^{T - 1} {\left| {l_{d,k}^{\left( i,I_2 \right)}} \right|}$, $i=\{0,1, \ldots, \tau-1\}$.
            \item [2)]\textbf{If} $\bar L_d^{\left( {i,I_2} \right)} < {\theta }$ for any $i \in \left\{ {0,1, \ldots ,\tau  - 1} \right\}$
                     \textbf{and} $w < w_{max}$  \\
                         (1) $w=w+1$; \\
                         (2) $I_{count} = 0$. \\
                         (3) Initialize the decoder with the channel LLRs of $w$ blocks.
          \end{itemize}
        \end{itemize}
\end{itemize}
\noindent\rule[.0cm]{0.5\textwidth}{0.6pt}

\begin{figure}[ht]
    \center
    \includegraphics[width= 0.4 \textwidth]{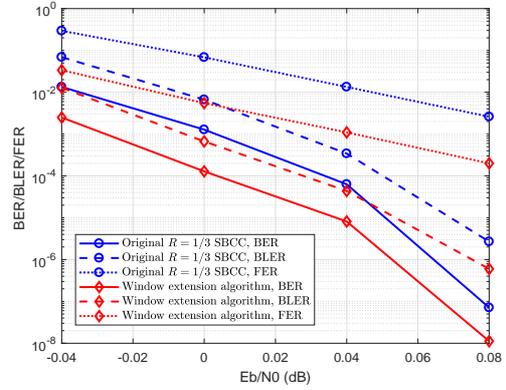}\\
    \caption{BER/BLER/FER performance comparison of rate $R=1/3$ blockwise SBCCs with and without the window extension algorithm.}
    \label{fig:BER_WinExt}
\end{figure}

For the same simulation conditions used in Fig. \ref{fig:Origin}, the BER, BLER, and FER performance of rate $R=1/3$ blockwise SBCCs with the window extension algorithm is shown in Fig. \ref{fig:BER_WinExt}, where $w_{max} = 6$, $\tau = 2$, and $\theta = 10$. We see that rate $R=1/3$ blockwise SBCCs with window extension show an order of magnitude improvement in BER, BLER, and FER compared to the results of Fig. \ref{fig:Origin}. We also remark that, even though $w_{max} = 6$, the average window size is only slightly larger $w$, e.g. $\bar w = 3.0014$ for ${E_b}/{N_0} = 0.04$ dB, since window extension is only employed when error propagation is detected.

\subsection{Resynchronization Mechanism}
We see from Fig. \ref{fig:BER_WinExt} that the window extension algorithm greatly reduces the effect of error propagation. However, for very long frames or for streaming, even one occurrence of error propagation can be catastrophic. We now introduce a resynchronization mechanism to address this problem.

As noted above, the first block in a BCC encoder chain has two known input sequences. Therefore, the input LLRs in the first block are more reliable than for the succeeding blocks. Motivated by this observation, and assuming the availability of a noiseless binary feedback channel, we propose that, when the window extension algorithm is unable to stop error propagation, the encoder resets to the initial $0$ state and begins the encoding of a new chain. This resynchronization mechanism is described below.

After a target block is decoded in the window extension algorithm, if its average absolute LLRs satisfy

\begin{equation}\label{eq:4}
\bar L_d^{\left( 0,I_2 \right)} < {\theta},
\end{equation}
we consider this target block as \emph{failed}. If we experience $N_r$ consecutive failed target blocks, we declare an error propagation condition and initiate encoder and decoder resychronization using the feedback channel. In other words, the encoder 1) sets the initial register states of the two component convolutional encoders to ``0'', and 2) begins encoding the next block with two known (all ``0'') input sequences together with the new information block. Meanwhile, the decoder makes decisions based on the current LLRs for the remaining blocks in the current window and restarts decoding once $w$ new blocks are received. Alg. 2 gives the detailed description.
%Full details of the resynchronization algorithm are given in an expanded version of the paper posted online [????].

%\begin{algorithm}[!t]
%\caption{Resynchronization Algorithm } \label{Alg:Reset_Alg}
%\begin{algorithmic}[1]
%\STATE Let $I_{count}$ and $R_{count}$ denote the current number of horizontal iterations and the counter for the average absolute LLRs of the target \hl{blocks}, respectively. Let $w$ denote the current window size\hl{, and set $I_{count}=0$ and $R_{count}=0$ initially.}
%\WHILE{$I_{count} < I_2$}
%\STATE \hl{Perform} vertical decoding and horizontal decoding;
%\STATE $I_{count} ++$;
%\ENDWHILE
%\STATE Calculate the average absolute LLRs \hl{$\bar L_d^{\left( 0,I_2 \right)}$ of the target block,}
%\IF {\hl{$\bar L_d^{\left( 0,I_2 \right)} < {\theta}$}}
% \STATE $R_{count}++$.
%\ELSE
% \STATE $R_{count}=0$.
%\ENDIF
%\IF {$R_{count} == N_r$}
%\STATE Resynchronize the encoder and decoder: 1) the initial register state of \hl{each component convolutional encoder} is ``0''; 2)\hl{ one input sequence of each component encoder is all ``0''}, the other input sequence is \hl{the new} information \hl{block}; 3) \hl{the decoder makes decisions based on the current LLRs for the remaining} blocks in the window and restarts decoding \hl{once} $w$ new blocks are received.
%\ENDIF
%
%\end{algorithmic}
%\end{algorithm}

\noindent\rule[0.0\baselineskip]{0.5\textwidth}{0.6pt}\\
$\textbf{~~~~~~~~Algorithm 2: Resynchronization Algorithm}$\\
\begin{itemize}
  \item Let $I_{count}$ and $R_{count}$ denote the current number of horizontal iterations and the counter for the average absolute LLRs of the target blocks, respectively. Let $w$ denote the current window size, and set $I_{count}=0$ and $R_{count}=0$ initially.
  \item \textbf{While} $I_{count} < I_2$ \\
         1) Perform vertical decoding and horizontal decoding; \\
         2) $I_{count} ++$;
  \item Calculate the average absolute LLRs $\bar L_d^{\left( 0,I_2 \right)}$ of the target block,
  \item \textbf{If} $\bar L_d^{\left( 0,I_2 \right)} < {\theta}$\\
         $R_{count}++$;\\
         \textbf{else}
         $R_{count}=0$.
  \item \textbf{If} $R_{count} == N_r$ \\
        Resynchronize the encoder and decoder: 1) the initial register state of each component convolutional encoder is ``0''; 2) one input sequence of each component encoder is all ``0'', the other input sequence is the new information block; 3) the decoder makes decisions based on the current LLRs for the remaining blocks in the current window and restarts decoding once $w$ new blocks are received.
\end{itemize}
\noindent\rule[.0cm]{0.5\textwidth}{0.6pt}

\begin{figure}[ht]
    \center
    \includegraphics[width= 0.4 \textwidth]{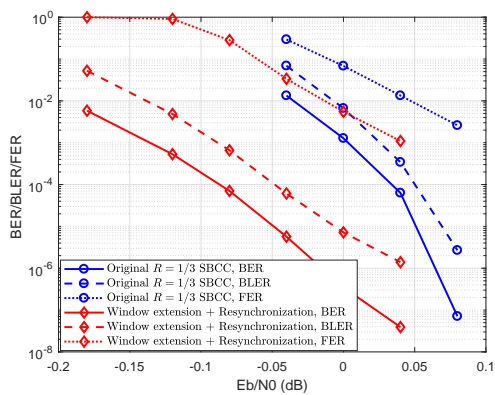}\\
    \caption{BER/BLER/FER comparison of rate $R=1/3$ blockwise SBCCs with and without window extension and resynchronization.}
    \label{fig:BER_WinExtReset}
\end{figure}

To demonstrate the efficiency of the resynchronization mechanism, the BER, BLER, and FER performance of rate $R=1/3$ blockwise SBCCs with the window extension algorithm and the resynchronization mechanism is shown in Fig. \ref{fig:BER_WinExtReset} for the same simulation conditions used in Fig. \ref{fig:BER_WinExt} and $N_r = 1$. We see that, compared to the results of Fig. \ref{fig:Origin}, rate $R=1/3$ blockwise SBCCs with window extension and resynchronization gain approximately three orders of magnitude in BER and BLER and about one order of magnitude in FER.\footnote{Although the resynchronization mechanism terminates error propagation in a frame, thus improving both BER and BLER, it does not further reduce the number of frames in error.}

\section{Early Stopping Rule}

The decoding complexity of BCCs with sliding window decoding depends mainly on the number of horizontal iterations. Therefore, in order to minimize unnecessary horizontal iterations, we introduce a soft BER stopping rule, which was first proposed for LDPC convolutional codes in \cite{Najeeb2016TCOM}.
Every time a horizontal iteration finishes, the average estimated BER $BE{R_{est}}$ of the target bits in the current window is obtained using the following steps.
\begin{itemize}
  \item Calculate the decision LLR (the sum of the channel LLR, the prior LLR, and the extrinsic LLR) $L_d^{j}$ of every information bit in the target block, $j=0,1,\ldots, T-1$;
  \item The average estimated BER of the target information bits is then given by $$BE{R_{est}} = \frac{1}{T}\sum\limits_{j = 0}^{T - 1} {1.0/\left( {1.0 + \exp \left( {\left| {L_d^j} \right|} \right)} \right)}.$$
\end{itemize}

If the average estimated BER of the target bits satisfies

\begin{equation}\label{eq:5}
{BER}_{est} \le \gamma,
\end{equation}
decoding is stopped and a decision on the target symbols in the current window is made.

\begin{figure}[ht]
        \center
    \includegraphics[width= 0.4 \textwidth]{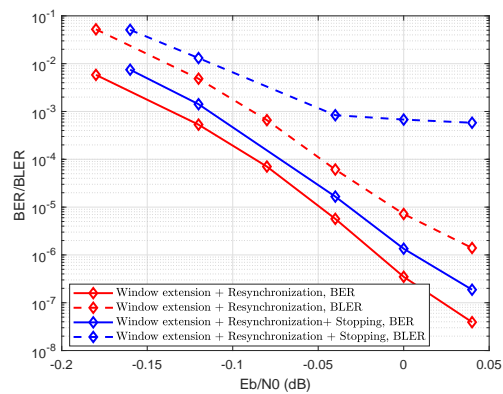}\\
    \caption{BER/BLER comparison of rate $R=1/3$ blockwise SBCCs with window extension and resynchronization, with and without the stopping rule.}
    \label{fig:BER_WinExtResetStop}
\end{figure}

\begin{figure}[ht]
    \center
    \includegraphics[width= 0.4 \textwidth]{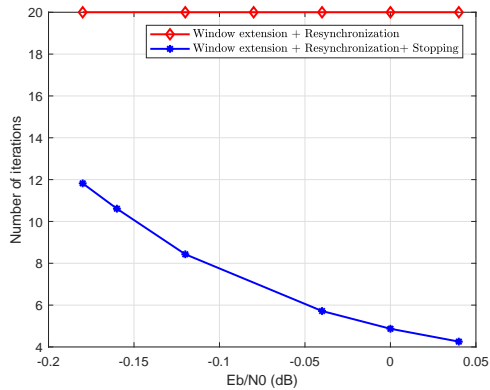}\\
    \caption{Number of horizontal iterations of rate $R=1/3$ blockwise SBCCs with window extension and resynchronization, with and without the stopping rule.}
    \label{fig:ITER_WinExtResetStop}
\end{figure}

Note that the window extension algorithm, the resynchronization mechanism, and the soft BER stopping rule can operate together in a sliding window decoder.
We now give an example to illustrate the tradeoffs between performance and computational complexity when error propagation mitigation is combined with the stopping rule. Fig. \ref{fig:BER_WinExtResetStop} shows the performance of rate $R=1/3$ blockwise SBCCs with window extension, resynchronization, and the stopping rule for the same simulation conditions used in Fig. \ref{fig:BER_WinExtReset} and $\gamma = 10^{-7}$. We see that using the stopping rule degrades the BER performance only slightly, but the BLER performance is negatively affected in the high SNR region.\footnote{The BLER loss at high SNR can be reduced by using a smaller $\gamma$ at a cost of some increased complexity.} The average number of horizontal iterations per block is also shown in Fig. \ref{fig:ITER_WinExtResetStop}, where we see that the stopping rule greatly reduces the number of horizontal iterations, especially in the high SNR region.

\section{Conclusion}

In this paper we investigated the error propagation problem associated with blockwise BCCs. A window extension algorithm and a resynchronization mechanism were proposed to mitigate error propagation, which can have a catastrophic effect on the performance for large frame lengths and continuous streaming operation. The BER and BLER performance of blockwise SBCCs with these two mitigation methods was shown to outperform the original blockwise SBCCs by about three orders of magnitude. Furthermore, a soft BER stopping rule was introduced and shown to significantly  reduce decoding complexity with little effect on BER performance.

%%%%%%
%% Appendix:
%% If needed a single appendix is created by
%%
%\appendix
%%
%% If several appendices are needed, then the command
%%
% \appendices
%%
%% in combination with further \section-commands can be used.
%%%%%%

%\section*{Acknowledgment}
%
%We are indebted to Michael Shell for maintaining and improving
%\texttt{IEEEtran.cls}.

%%%%%%
%% To balance the columns at the last page of the paper use this
%% command:
%%
%\enlargethispage{-1.2cm}
%%
%% If the balancing should occur in the middle of the references, use
%% the following trigger:
%%
%%\IEEEtriggeratref{3}
%%
%% which triggers a \newpage (i.e., new column) just before the given
%% reference number. Note that you need to adapt this if you modify
%% the paper.  The "triggered" command can be changed if desired:
%%
%\IEEEtriggercmd{\enlargethispage{-20cm}}
%%
%%%%%%

%%%%%%
%% References:
%% We recommend the usage of BibTeX:
%%
%\bibliographystyle{IEEEtran}
%\bibliography{definitions,bibliofile}

\begin{thebibliography}{9}

%\bibitem{WeizhangISIT2005}
%W.~Zhang, M.~Lentmaier, K.~Sh.~Zigangirov, and D.~J.~Costello,~Jr., ``Braided convolutional codes,'' in \emph{Proc. IEEE Int. Symp. Information Theory},~Adelaide, Australia, Sep. 2005, pp. 592-596.

\bibitem{WeizhangIT2010}
W.~Zhang, M.~Lentmaier, K.~Sh.~Zigangirov, and D.~J.~Costello,~Jr., ``Braided convolutional codes: a new class of turbo-like codes,'' \textit{IEEE Trans. Inf. Theory}, vol.~56, no.~1, pp.~316-331, Jan. 2010.

%\bibitem{Truhachev2003ISIT}
%D. Truhachev, M. Lentmaier, and K. Zigangirov, ``On braided block codes,'' in \textit{Proc. IEEE International Symposium on Informaiton Theory} (ISIT), Yokohama, Japan, 29 June-4 July 2003, pp. 32.

\bibitem{FeltstromIT2009}
A. J. Feltstr\"{o}m, M.~Lentmaier, D. V. Truhachev, and K. S. Zigangirov, ``Braided block codes,'' \textit{IEEE Trans. Inf. Theory}, vol.~55, no.~6, pp.~2640-2658, Jun. 2009.

\bibitem{Elias1954}
P. Elias, ``Error free coding,'' \textit{IRE Trans. Inf. Theory}, vol.~4, no.~4, pp.~29-37, Sep. 1954.

\bibitem{Sipser1996}
M. Slipser and D. A. Spielman, ``Expander codes,'' \textit{IEEE Trans. Inf. Theory}, vol.~42, no.~6, pp.~1710-1722, Nov. 1996.

%\bibitem{Zemor2001}
%G. Z\'{e}mor, ``On expander codes,'' \textit{IEEE Trans. Inf. Theory}, vol.~47, no.~2, pp.~835-837, Feb. 2001.
%
%\bibitem{Costell2016ISTC}
%D. J. Costello, Jr., M. Lentmaier, and D. G. M. Mitchell, ``New perspective on braided convolutional codes,'' in \textit{Proc. 9th International Symposium on Turbo Codes and Iterative Information Processing} (ISTC), Brest, France, Sept. 2016, pp. 2165-4719.

\bibitem{Saeedeh2017IT}
S. Moloudi, M. Lentmaier, and A. Graell i Amat, ``Spatilly coupled turbo-like codes,'' \textit{IEEE Trans. Inf. Theory}, vol.~63, no. 10, pp. 6199-6215, 2017.

%\bibitem{Saeedeh2014ISIT}
%S. Moloudi and M. Lentmaier, ``Density evolution analysis of braided convolutional codes on the erasure channel,'' in \emph{Proc. IEEE Int. Symp. Information Theory}, Honolulu, HI, USA, July 2014, pp. 2609-2613.
%
%\bibitem{Saeedeh2015ITW}
%S. Moloudi, M. Lentmaier, and A. Graell i Amat, ``Threshold saturation for spatially coupled turbo-like codes over the binary erasure channel,'' in \emph{Proc. IEEE Information Theory Workshop - Fall (ITW)}, Jeju Island, Korea, Oct. 2015, pp. 138-142.
%
\bibitem{Saeedeh2016ISITA}
S. Moloudi, M. Lentmaier, and A. Graell i Amat, ``Finite length weight enumerator analysis of braided convolutional codes,'' in \emph{Proc. International Symposium on Information Theory and Its Applications}, Monterey, CA, USA, Oct. 2016, pp. 488-492.
%
%\bibitem{LentmaierISIT2005}
%M.~Lentmaier, A.~Sridharan, K.~S.~Zigangirov, and D.~J.~Costello,~Jr., ``Terminated LDPC convolutional codes with thresholds close to capcacity,'' in \emph{Proc. IEEE Int. Symp. Information Theory},~Adelaide, Australia, Sep. 2005, pp. 1372-1376.

\bibitem{LentmaierTIT2010}
M.~Lentmaier, A.~Sridharan, D.~J.~Costello,~Jr., and K.~S.~Zigangirov, ``Iterative decoding threshold analysis for LDPC convolutional codes,'' \textit{IEEE Trans. Inf. Theory}, vol. 56, no. 10, pp. 5274-5289, Oct. 2010.

\bibitem{IyengarTIT2012}
A.~R.~Iyengar, M.~Papaleo, P.~H.~Siegel, J.~K.~Wolf, A.~Vanelli-Coralli, and G.~E.~Corazza, ``Windowed decoding of protograph-based LDPC convolutional codes over erasure channels,'' \textit{IEEE Trans. Inf. Theory}, vol. 58, no. 4, pp. 2303-2320, April 2012.

%\bibitem{Zhu2015ITW}
%M. Zhu, D. G. M. Mitchell, M. Lentmaier, D. J. Costello, Jr., and B. Bai, ``Window decoding of braided convolutional codes,'' in \emph{Proc. IEEE Information Theory Workshop - Fall (ITW)}, Jeju Island, Korea, Oct. 2015, pp. 143-147.

\bibitem{Zhu2017TCOM}
M. Zhu, D. G. M. Mitchell, M. Lentmaier, D. J. Costello, Jr., and B. Bai, ``Braided convolutional codes with sliding window decoding,'' \textit{IEEE Trans. on Communications}, vol. 65, no. 9, pp. 3645-3658, Sept. 2017.

\bibitem{Najeeb2016TCOM}
N. Ul Hassan, A E. Pusane, M. Lentmaier, G. P. Fettweis, and D. J. Costello, Jr., ``Non-uniform window decoding schedules for spatially coupled LDPC codes,'' \textit{IEEE Trans. on Communications}, vol. 65, no. 2, pp. 501-510, Nov. 2016.

\end{thebibliography}
%%
%% where we here have assume the existence of the files
%% definitions.bib and bibliofile.bib.
%% BibTeX documentation can be obtained at:
%% http://www.ctan.org/tex-archive/biblio/bibtex/contrib/doc/
%%%%%%

%% Or you use manual references (pay attention to consistency and the
%% formatting style!):

\end{document}